\documentclass[12pt]{article}  
\begin{document}

\newcommand{\sheptitle}
{ Right-handed Majorana Neutrino Mass Matrices for Generating Bimaximal Mixings in Degenerate and Inverted Models of Neutrinos}
\newcommand{\shepauthor}
{Mahadev Patgiri${^\dag}$\footnote{corresponding author.\\ {\it{e-mail address:}} mahadev@scientist.com}   and N. Nimai Singh${^\ddag}$ }

\newcommand{\shepaddress}
{${^\dag}$Department of Physics, Cotton College, Guwahati-781001, India\\
${^\ddag}$ Department of Physics, Gauhati University, Guwahati-781014, India}

\newcommand{\shepabstract}
{An attempt is made to generate the bimaximal mixings of the three species of neutrinos 
from the textures of the right-handed Majorana neutrino mass matrices.  We extend our earlier work 
 in this paper for the
generation of  the nearly degenerate as well as the inverted
hierarchical models of the left-handed Majorana neutrino mass matrices 
using the non-diagonal textures of the right-handed Majorana neutrino mass matrices
and  the diagonal form of Dirac neutrino mass matrices, 
within the framework of the see-saw mechanism  in a model independent way.
 Such  Majorana neutrino mass models 
are important to explain  the recently reported  result on 
the neutrinoless double beta decay $(0\nu\beta\beta)$ experiment, together with the earlier 
established data on LMA MSW solar and atmospheric neutrino oscillations.\\

PACS numbers: 12.10.Dm;12.15.Ff;14.60.Gh\\

Keywords: Majorana neutrino mass, degenerate, inverted hierarchical, mixing angles, mass eigenvalues.}
\begin{titlepage}
\begin{flushright}
hep-ph/0301254
\end{flushright}
\begin{center}
{\large{\bf\sheptitle}}
\bigskip\\
\shepauthor
\\
\mbox{}\\
{\it\shepaddress}\\
\vspace{.5in}
{\bf Abstract}
\bigskip
\end{center}
\setcounter{page}{0}
\shepabstract
\end{titlepage}
\section{Introduction}
In the context of the recently reported experimental result on double beta decay[1], together with 
the earlier experimental data on the atmospheric[2] and solar[3] neutrino oscillations,
 it is important to construct theoretical models which predict the degenerate and inverted 
hierarchical patterns of the Majorana 
neutrino mass matrices
within  the framework of the grand unified theories(GUTs) with or without 
supersymmetry[4,5]. In this short paper we attempt
to generate the degenerate as well as the inverted hierarchical pattern of the left-handed 
Majorana neutrino mass matrices using the see-saw formula in a model independent way.
This is, in fact, a continuation of our earlier work[6]  where the neutrino mixings 
are provided from the texture of the right-handed Majorana mass matrix $M_{RR}$,
while keeping the Dirac neutrino mass matrix $m_{LR}$ in the diagonal form. 
We had taken the Dirac neutrino mass matrix $m_{LR}$ as either the charged lepton mass matrix
( $m_{LR}$ =$ tan\beta$ $m_{l}$ referred to as case(i)) or the up-quark mass matrix
( $m_{LR} = m_{up}$ referred to as case(ii))[7]. While referring to the earlier paper[6]
for details, the model successfully generated both the hierarchical and the inverted
hierarchical (having opposite sign mass eigenvalues) neutrino mass matrices as a
result of the proper choice of the parameters in texture of $M_{RR}$. In section 2 we present 
the generation of the degenerate as well as inverted hierarchical neutrino mass matrices using 
the see-saw formula, and their predictions on mass eigenvalues and mixing angles. Section 3 is devoted to 
summary and conclusion.

\section{Neutrino mass matrices from see-saw formula}

The left-handed Majorana neutrino mass matrix $m_{LL}$ is
given by the celebrated see-saw formula[8],
\begin{equation}
m_{LL} = -m_{LR} M^{-1}_{RR} m^{T}_{LR}
\end{equation}
where  $m_{LR}$ is the Dirac neutrino mass matrix in the left-right (LR) convention[9]. 
The leptonic (MNS) mixing matrix is now given by $V_{MNS} = V^{\dag}_{\nu L}$
where $m^{diag}_{LL}$ = $V_{\nu L}$ $ m_{LL}$ $V^{T}_{\nu L}$. Here both  $m_{LR}$
and the charged lepton mass matrix $m_{l}$ are taken as diagonal, whereas the
right-handed Majorana neutrino mass matrix $M_{RR}$ as non-diagonal. Using 
the see-saw formula (1) we generate both patterns of $m_{LL}$ viz.,
(I) nearly degenerate and (II) inverted hierarchical neutrino mass models. We conentrate here only on 
the cases which have  bimaximal mixings  listed in Table-I.\\

Table-I: Zeroth order neutrino mass matrices with texture zeros 
corresponding to the LMA MSW solution with bimaximal mixings [4,10].

\begin{center}
\begin{tabular}{ccc} \hline \hline
\it{Type} & \it{$m_{LL}$} & \it{$m^{diag}_{LL}$}\\
\hline
I(A) & $\left( \begin{array}{ccc}
 0 & \frac{1}{\sqrt{2}}& \frac{1}{\sqrt{2}}  \\
  \frac{1}{\sqrt{2}}  & \frac{1}{2} & -\frac{1}{2}\\
  \frac{1}{\sqrt{2}} & -\frac{1}{2}& \frac{1}{2}
\end{array}
\right) m_{0}$ & $Diag( 1, -1, 1) m_{0}$\\

I(B) & $\left( \begin{array}{ccc}
 1 & 0 &  0\\
 0 & 1 & 0\\
 0 & 0 & 1
\end{array}
\right) m_{0}$ & $Diag( 1, 1, 1) m_{0}$\\
I(C) & $\left( \begin{array}{ccc}
 1 & 0 &  0\\
 0 & 0 &   1\\
 0&  1 & 0
\end{array}
\right) m_{0}$ & $Diag( 1, 1, -1) m_{0}$\\

II(A) & $\left( \begin{array}{ccc}
 1 & 0 &  0\\
 0 & \frac{1}{2}& \frac{1}{2}\\
 0& \frac{1}{2} & \frac{1}{2}
\end{array}
\right) m_{0}$ & $Diag( 1, 1, 0) m_{0}$\\

II(B) &  $\left( \begin{array}{ccc}
 0 & 1 &  1\\
 1 & 0 & 0\\
 1 & 0 & 0
\end{array}
\right) m_{0}$ & $Diag( 1, -1, 0) m_{0}$\\
\hline\hline
\end{tabular}
\end{center}

The Dirac neutrino mass matrix $m_{LR}$ involved in the see-saw formula Eq.(1),
can be either the charged lepton mass matrix $m_{l}$ (case (i))or the up-quark 
mass matrix $m_{up}$ (case (ii)) depending on the particular SUSY SO(10) GUT model 
and the contents of the Higgs fields employed[6,7]:
\begin{equation}
m_{LR}= tan\beta
\left( \begin{array}{ccc}
\lambda^{6} & 0 & 0 \\
0 & \lambda^{2} & 0 \\
0 & 0 & 1
\end{array}
\right) m_{\tau},
\end{equation}
and

\begin{equation}
m_{LR}= 
\left( \begin{array}{ccc}
\lambda^{8} & 0 & 0 \\
0 & \lambda^{4} & 0 \\
0 & 0 & 1
\end{array}
\right) m_{t},
\end{equation}
respectively. The above two forms of $m_{LR}$ may be written together as 
\begin{equation}
m_{LR}= 
\left( \begin{array}{ccc}
\lambda^{m} & 0 & 0 \\
0 & \lambda^{n} & 0 \\
0 & 0 & 1
\end{array}
\right) m_{f},
\end{equation}
where $m_{f}$ corresponds to $(m_{\tau}\tan\beta)$ in SUSY models for charged lepton mass matrix in case (i), and $m_{t}$ for up-quark mass matrix in case (ii). The pair of the exponents $(m,n)$ are $(6,2)$ for charged lepton and $(8,4)$
 for up-quark mass matrices respectively. The value of the Wolfenstein parameter is taken as $\lambda=0.22$.
Using the diagonal form of $m_{LR}$ in Eq.(4) and a suitable choice of the non-diagonal texture of $M_{RR}$, 
the following four types of neutrino mass matrices $m_{LL}$ are calculated.

{\bf I(A). Nearly degenerate mass matrix with opposite sign mass eigenvalues}\\
The degenerate mass matrix $m_{LL}$ having opposite sign mass eigenvalues  is now 
generated through see-saw formula[8] in Eq.(1)
for the  choices of $m_{LR}$ in Eq.(4) and 
\begin{equation}
M_{RR} =
\left( \begin{array}{ccc}
-2\delta_{2}\lambda^{2m} & (\frac{1}{\sqrt2}+\delta_{1})\lambda^{m+n} & (\frac{1}{\sqrt2}+\delta_{1})\lambda^{m}\\
(\frac{1}{\sqrt2}+\delta_{1})\lambda^{m+n} & (\frac{1}{2}+\delta_{1}-\delta_{2})\lambda^{2n} & (-\frac{1}{2}+\delta_{1}-\delta_{2})\lambda^{n}\\
(\frac{1}{\sqrt2}+\delta_{1})\lambda^{m} &  (-\frac{1}{2}+\delta_{1}-\delta_{2})\lambda^{n} & (\frac{1}{2}+\delta_{1}-\delta_{2})
\end{array}
\right) v_{R}
\end{equation}    
leading to a simple form,
\begin{equation}
m_{LL}= 
\left( \begin{array}{ccc}
 -2\delta_{1}+2\delta_{2}& \frac{1}{\sqrt{2}}-\delta_{1}& \frac{1}{\sqrt{2}}-\delta_{1}  \\
  \frac{1}{\sqrt{2}}-\delta_{1}  & \frac{1}{2}+\delta_{2} & -\frac{1}{2}+\delta_{2}\\
  \frac{1}{\sqrt{2}}-\delta_{1} & -\frac{1}{2}+\delta_{2} & \frac{1}{2}+\delta_{2}
\end{array}
\right) m_{0}
\end{equation}
where $m_{0}$ controls the overall magnitude of the masses of the 
neutrinos whereas $\delta_{1}$ and $\delta_{2}$ give the desired
splittings for solar and atmospheric data. When $\delta_{1}=\delta_{2}= 0$, 
Eq.(6) reduces to the zeroth order mass matrix of  
the Type I(A) in Table-I, with no splittings[10].
The diagonalisation of $m_{LL}$ in Eq.(6) leads to the following eigenvalues and mixings:\\
$m_{\nu_{1}}= [1+2\delta_{2}-\delta_{1}(1+\sqrt{2})]m_{0},$\\
$m_{\nu_{2}}=[-1+2\delta_{2}-\delta_{1}(1-\sqrt{2})]m_{0},$\\
$m_{\nu_{3}}=m_{0}$,\\
$\sin^{2}2\theta_{12}\approx (1-\delta_{1}^{2}/8)$, $\sin^{2}2\theta_{23}= 1$,
$\sin^{2}2\theta_{13}=0$
\\
For the choice  of the values of the parameters $m_{0}=0.4$ eV,
$\delta_{1}=0.0061875$, $\delta_{2}=0.0030625$,
Eq.(6) leads to the following numerical  predictions \\
\underline{\it{Mixing angles}}:\\
$sin^{2}2\theta_{12}= 0.999$, $sin^{2}2\theta_{23} \approx 1.0$,
$|V_{e3}| = 6.124\times 10^{-9}$,\\
\underline{\it{Mass eigenvalues}}:\\
$m_{\nu_{i}} = (0.396484, -0.396532, 0.4)$ eV, $i = 1, 2, 3$;
$\Delta m^{2}_{12} = 3.806\times  10^{-5} eV^{2}$ and
$\Delta m^{2}_{23} = 2.76\times 10^{-3} eV^{2}$.\\
The prediction on solar mixing angle is maximal and larger than the  LMA MSW solution.
The expression for $m_{0}$ in Eq.(6) for case (i) is worked out as
$m_{0} = m^{2}_{\tau} tan^{2}{\beta}/ v_{R}$. 
For input values of $m_{0} = 0.4 eV$, $tan\beta = 40$, $m_{\tau} = 1.7$ GeV, 
the see-saw scale is calculated as $v_{R} \approx 10^{13}$ GeV.  
This in turn  gives the masses of the three right-handed Majorana neutrinos after the 
diagonalisation of $M_{RR}$:\\
$|M^{diag}_{RR}| = ( 5.0427\times 10^{12}, 3.0981\times10^{8}, 1.9613\times 10^{7})$ GeV.\\
Similarly, in  case(ii) we  have $m_{0} = m^{2}_{t}/ v_{R}$ in Eq.(6), and with the input values $m_{0} = 0.4$ eV,
$m_{t} = 200$ GeV, we obtain $v_{R}= 10^{14}$ GeV and the mass eigenvalues of the right-handed
Majorana neutrinos: 
$|M^{diag}_{RR}| = (4.5932\times 10^{15}, 7.2731\times 10^{6}, 5.005 \times 10^{13})$ GeV.\\

{\bf{I(B). Nearly degenerate mass matrix with the same sign mass eigenvalues}}\\
The  mass matrix $m_{LL}$ of this type can be realised
in the see-saw mechanism(1) using the general  texture of $m_{LR}$ in Eq.(4) and 
\begin{equation}
M_{RR} =
\left( \begin{array}{ccc}
(1+2\delta_{1}+2\delta_{2})\lambda^{2m} & \delta_{1}\lambda^{m+n} &  \delta_{1}\lambda^{m}\\
 \delta_{1}\lambda^{m+n} &  (1+\delta_{2})\lambda^{2n} &  \delta_{2}\lambda^{n}\\
 \delta_{1}\lambda^{m} &  \delta_{2}\lambda^{n} & (1+\delta_{2})
\end{array}
\right) v_{R}
\end{equation}
leading to  the nearly degenerate mass matrix,
\begin{equation}
m_{LL}=
\left( \begin{array}{ccc}
(1-2\delta_{1}-2\delta_{2}) & -\delta_{1} &  -\delta_{1}\\
 -\delta_{1} &(1 -\delta_{2}) & -\delta_{2}\\
 -\delta_{1}  & -\delta_{2} &  (1-\delta_{2})
\end{array}
\right) m_{0}
\end{equation}
The diagonalisation of $m_{LL}$ in Eq.(8) leads to \\
$m_{\nu_{1}}\simeq (1- 2\delta_{2}- (\sqrt{3}+1)\delta_{1})m_{0},$\\
$m_{\nu_{2}}\simeq (1- 2\delta_{2}+ (\sqrt{3}-1)\delta_{1})m_{0},$\\
$m_{\nu_{3}}\simeq m_{0}$,\\
$\sin^{2}2\theta_{12}= \frac{2}{3}$, $\sin^{2}2\theta_{23}= 1$,
$\sin^{2}2\theta_{13}= 0$.\\
For the choice  of the values of the parameters $m_{0}=0.4$ eV,
$\delta_{1}=3.6 \times 10^{-5}$, $\delta_{2}=3.9 \times 10^{-3}$,
Eq.(8) leads to the following numerical  predictions: \\
\underline{\it{Mixing angles}}:\\
$sin^{2}2\theta_{12}= 0.67$, $sin^{2}2\theta_{23} \approx 1.0$,
$|V_{e3}| = 1.5 \times 10^{-14}$,\\
\underline{\it{Mass eigenvalues}}:\\
$m_{\nu_{i}} = ( 0.39684, 0.396892, 0.4)$ eV, $i = 1, 2, 3$;
$\Delta m^{2}_{12} = 4.13 \times  10^{-5} eV^{2}$ and
$\Delta m^{2}_{23} = 2.48 \times 10^{-3} eV^{2}$.\\
The prediction on solar mixing angle  is consistent with the LMA MSW solution[3].\\
For case(i), the expression for $m_{0}$ in Eq.(8) is again worked out as
$m_{0} = m^{2}_{\tau} tan^{2}{\beta}/ v_{R}$, and 
for input values of $m_{0}=0.4eV$, $\tan\beta=40$, $m_{\tau}=1.7$GeV,
we obtain $v_{R}=1.156\times10^{13}$GeV which leads to
$|M_{R}^{diag}|=(1.15\times10^{13},2.71\times10^{10},1.498\times10^{5})$GeV.
Similarly for case (ii), we have  $m_{0} = m^{2}_{t}/ v_{R}$ and 
for input values of $m_{0}=0.4eV$ and $m_{t}=200$GeV, we have $v_{R}=1.0\times10^{14}$GeV 
and $|M^{diag}_{RR}|=(1.0039\times10^{14}, 5.5089\times10^{8}, 3.035\times10^{4})$GeV.\\

{\bf{I(C). Nearly degenerate mass matrix with opposite sign mass eigenvalues}}\\ 
We consider another texture for the nearly degenerate mass matrix  
$m_{LL}$ with opposite mass eigenvalues[5].  
We take $m_{LR}$ given in Eq.(4) and the following  right-handed 
neutrino mass matrix
\begin{equation}
M_{RR} =
\left( \begin{array}{ccc}
(1+2\delta_{1}+2\delta_{2})\lambda^{2m} & \delta_{1}\lambda^{m+n} &  \delta_{1}\lambda^{m}\\
\delta_{1}\lambda^{m+n} &  \delta_{2}\lambda^{2n} &  (1+\delta_{2})\lambda^{n}\\
\delta_{1}\lambda^{m} &  (1+\delta_{2})\lambda^{n} & \delta_{2}
\end{array}
\right) v_{R}
\end{equation}    
The resulting $m_{LL}$ is calculated as 
\begin{equation}  
m_{LL}= 
\left( \begin{array}{ccc} 
(1-2\delta_{1}-2\delta_{2}) & -\delta_{1} &  -\delta_{1}\\ 
-\delta_{1} & -\delta_{2} & (1-\delta_{2})\\ 
-\delta_{1}  & (1-\delta_{2}) &  -\delta_{2} 
\end{array} 
\right) m_{0} 
\end{equation} 
where $m_{0}$ controls the overall magnitude of the masses of the  
neutrinos whereas $\delta_{1}$ and $\delta_{2}$ give the desired 
splittings for solar and atmospheric data. When $\delta_{1}=\delta_{2}= 0$,  
Eq.(10) reduces to the zeroth order mass matrix of   
the Type I(C) in Table-I, with no splittings[5,10]. 
The diagonalisation of $m_{LL}$ in Eq.(10) leads to the following 
eigenvalues and mixing angles:\\ 
$m_{\nu_{1}}\simeq (1- 2\delta_{2}- (\sqrt{3}+1)\delta_{1})m_{0},$\\
$m_{\nu_{2}}\simeq (1- 2\delta_{2}+ (\sqrt{3}-1)\delta_{1})m_{0},$\\
$m_{\nu_{3}}\simeq -m_{0},$\\
$\sin^{2}2\theta_{12}= \frac{2}{3}$, $\sin^{2}2\theta_{23}= 1$,
$\sin^{2}2\theta_{13}= 0.$
\\
The numerical solution leads to   
$m_{\nu_{i}} = (0.39684, 0.396892, -0.4)$ eV, $i=1, 2, 3$
for the same choices of the input values of $\delta_{1,2}$ and $m_{0}$
as in Eq.(8). Further, the predictions on the three mixing angles
are the same as in Eq.(8). 
When $\delta_{1}= \delta_{2}= 0$, it reduces to the Type I(C)
in the Table-I.
In case (i), for input values of $m_{0} = 0.4 eV$, $tan\beta = 40$, $m_{\tau} = 1.7$ GeV,
 the see-saw scale is calculated as $v_{R} \approx 10^{13}$ GeV.  
This in turn  gives the masses of the three right-handed Majorana neutrinos after the 
diagonalisation of $M_{RR}$:
$|M^{diag}_{RR}| = ( 4.67\times 10^{11}, 1.296\times10^{5}, 5.06\times 10^{4})$ GeV.
Similarly for case (ii), we have $m_{0} = m^{2}_{t}/ v_{R}$ in Eq.(10), and with the input 
values $m_{0} = 0.4$ eV,
$m_{t} = 200$ GeV, we obtain $v_{R}= 10^{14}$ GeV and the mass eigenvalues of the right-handed Majorana neutrinos: 
$|M^{diag}_{RR}| = ( 1.105 \times 10^{11}, 3.035 \times 10^{3}, 5.005 \times 10^{11})$ GeV.\\

{\bf II(A). Inverted hierarchical mass matrix with same sign mass eigenvalues}\\
The most general form of the inverted hierarchical mass matrix $m_{LL}$ 
with the same sign mass eigenvalues, can be calculated  with the choice of $m_{LR}$ 
given in Eq.(4) and $M_{RR}$ of the following form
\begin{equation}
M_{RR}=
\left( \begin{array} {ccc}
2a\eta (1+2\epsilon) \lambda^{2m} & \eta \epsilon \lambda^{m+n} & \eta \epsilon \lambda^{m}\\ 
\eta \epsilon \lambda^{m+n}  &  a\lambda^{2n}  &  -(a-\eta)\lambda^{n}\\
\eta \epsilon \lambda^{m}  &  -(a-\eta)\lambda^{n} & a
\end{array} \right) \frac{v_{R}}{2a\eta}
\end{equation}
leading to 
\begin{equation}
m_{LL}=
\left( \begin{array}{ccc}
(1-2\epsilon) & -\epsilon &  -\epsilon\\
 -\epsilon & a  & (a-\eta)\\
 -\epsilon  & (a-\eta) & a  
\end{array}
\right) m_{0}
\end{equation}
where $a =0.5$ and $m_{0}$ is the overall factor for 
the masses of the neutrinos.
The parameters $\epsilon$ and $\eta$ give the desired splittings for solar
and atmospheric data.
The diagonalisation of $m_{LL}$ in Eq.(12)
leads to the following eigenvalues and mixing angles:\\
$m_{\nu_{1}}\simeq (1- (\sqrt{3}+1)\epsilon-
\frac{\eta}{2}+\frac{\sqrt{\eta\epsilon}}{6})m_{0},$\\
$m_{\nu_{2}}\simeq (1+ (\sqrt{3}-1)\epsilon
-\frac{\eta}{2}-\frac{\sqrt{\eta\epsilon}}{6})m_{0},$\\
$m_{\nu_{3}}$$\simeq$ $\eta$ $m_{0},$\\
and mixing angles:\\
 $\sin^{2}2\theta_{12}= \frac{2}{3}$, 
$\sin^{2}2\theta_{23}= 1$,
$\sin^{2}2\theta_{13}= 0.$\\
When $\epsilon = \eta = 0$, Eq.(12) reduces to
the zeroth order mass matrix of the type II(A) in Table-I,
with no solar splitting[4,10].  
For solution of the 
LMA MSW solar data and atmospheric neutrino oscillation, we have the choice
of the parameters $m_{0} = 0.05$ eV, $\epsilon = 0.002$ and $\eta = 0.0001$
leading to the following predictions:\\
\underline{\it{Mixing angles}:}\\
 $sin^{2}2\theta_{12}= 0.67$, $sin^{2}2\theta_{23} \approx 1.0$,
$|V_{e3}| = 3.04 \times 10^{-13}$,\\
\underline{\it{Mass eigenvalues}}:\\
$m_{\nu_{i}} = ( 0.05007, 0.04973, 0.000005)$ eV, $i = 1, 2, 3$; leading to
$\Delta m^{2}_{12} = 3.393 \times  10^{-5} eV^{2}$ and 
$\Delta m^{2}_{23} = 2.47 \times 10^{-3} eV^{2}$.

The expression for $m_{0}$ in Eq.(12) for case (i)) is given by
$m_{0} = m^{2}_{\tau} tan^{2}\beta/v_{R}$. For input values of $m_{0}= 0.05$ eV, $tan\beta= 5$,
$m_{\tau}= 1.7$ GeV, we obtain $v_{R}= 1.445\times 10^{12}$ GeV which 
leads to $|M^{diag}_{RR}|= (9.742\times 10^{3}, 2.831
\times 10^{4}, 7.24\times10^{15})$ GeV.
Again for case(ii) $m_{0}= m^{2}_{t}/v_{R}$ in Eq.(12). Using the input values
$m_{0}=0.05 eV$, $m_{t}= 200$GeV, we have $v_{R}= 8\times 10^{14}$ GeV and
$|M^{diag}_{RR}|= ( 2.4\times 10^{4}, 4\times 10^{18}, 2.4\times 10^{9})$ GeV
where the mass of the heaviest right-handed Majorana neutrino lies above
the GUT scale but below the Planck scale[10].\\

{\bf II(B). Inverted hierarchical mass matrix with opposite  sign mass eigenvalues}\\ 
 
Here the first two neutrino mass eigenvalues are of opposite  sign and this 
 inverted hierarchical mass matrix has  the following form[11],
\begin{equation}
m_{LL}=
\left( \begin{array}{ccc}
\epsilon  & 1 &  1\\
 1 & \delta_{1}  & \delta_{2}\\
 1  & \delta_{2} & \delta_{1}  
\end{array}
\right) m_{0},  \epsilon,\delta_{1},\delta_{2}<< 1
\end{equation}
 For $ \delta_{1}$, $\delta_{2}$,
$\epsilon$ = 0, it leads to the type II(B) in Table-I. 
This structure has been successfully generated
within this model in our earlier paper[6], and without much details we outline one case only.
As an example, for case (ii) where $m_{LR}=m_{up}$, we have
\begin{equation}
m_{LL}=
\left( \begin{array}{ccc}
0  & 1 &  1\\
 1 & \lambda^{3}  & 0\\
 1  & 0 & \lambda^{3}  
\end{array}
\right)m_{0}
\end{equation}
for the texture of $M_{RR}$:
\begin{equation}
M_{RR}=
\left( \begin{array}{ccc}
-\lambda^{22}  & \lambda^{15} &  \lambda^{11}\\
 \lambda^{15} & \lambda^{8}  & -\lambda^{4}\\
 \lambda^{11}  & -\lambda^{4} & 1  
\end{array}
\right)v_{R}
\end{equation}
The predictions are\\
$m_{i}=(1.4195,-1.4089,0.0105)m_{0}$, $i=1,2,3$\\
$sin^{2}2\theta_{12}= 0.9999$, $sin^{2}2\theta_{23} \approx 1.0$,
$|V_{e3}| = 1.11 \times 10^{-16}$,\\
For input $m_{0}=0.05eV$ we get the correct mass eigenvalues and $v_{R}=4.0\times 10^{16}GeV$. This gives\\ 
$|M^{diag}_{RR}| = ( 3.055 \times 10^{-12}, 2.44 \times 10^{-8}, 1.0)v_{R}$ GeV.\\

The predictions of the solar mixing angle in all types except in types I(A) and II(B),
 agree with the LMA MSW solution.  The solar mixings predicted from $m_{LL}$ in Eqs.(6) and (14)
 (Types I(A) and II(B)) are above the upper experimental limit[3]
$\sin^{2}2\theta_{12}\leq 0.988$ with the best fit value  
$\sin^{2}2\theta_{12}= 0.8163$. Any fine tuning can hardly improve  $\sin^{2}2\theta_{12}$.
 One may expect some spectacular changes if the contribution from the diagonalisation of the 
charged lepton mass matrix having special entries in the 1-2 block, is taken into consideration  
in the MNS mixing matrix $V_{MNS}=V_{eL}V^{\dag}_{\nu L}$. For example, such  a  charged lepton mass 
matrix may have the following  structure[12,13] 
\begin{equation}
m_{l}=
\left( \begin{array}{ccc}
0.00256  & -0.01058 &  0\\
 -0.01058 & 0.04596  & 0\\
 0  & 0 & 1  
\end{array}
\right)m_{\tau}
\end{equation}
and  its diagonalisation leads to the following  diagonalisation matrix and eigenvalues:
\begin{equation}
V_{eL}=
\left( \begin{array}{ccc}
-0.97439  & -0.22488 &  0\\
 -0.22488 & 0.97439  & 0\\
 0  & 0 & 1  
\end{array}
\right)
\end{equation}
and $m^{diag}_{l}=(1.182\times 10^{-4}, 4.840\times 10^{-2}, 1.0)m_{\tau}$.
With the contribution from charged lepton to the MNS matrix $V_{MNS}=V_{eL}V^{\dag}_{\nu L}$, 
 the mixings angles are calculated  as \\
$sin^{2}2\theta_{12}= 0.8517$, $sin^{2}2\theta_{23}=0.9494$, and $|V_{e3}|=0.159$
for Type II(B) in Eq.(14), and 
$sin^{2}2\theta_{12}= 0.8576$, $sin^{2}2\theta_{23}=0.9495$, and $|V_{e3}|=0.159$
for Type I(A) in Eq.(6). These results are now consistent with the LMA MSW solution 
for solar neutrino anomaly and maximal mixing for the atmospheric neutrino oscillation, along 
with the CHOOZ constraint $|V_{e3}|\leq 0.16$. The corresponding left-handed neutrino mass 
matrix in the diagonal basis of the charged lepton mass matrix, is  given by relation 
$m'_{LL}=V_{eL}m_{LL}V^{\dag}_{eL}$. For Type II(B) in Eq.(14) we obtain
\begin{equation}
m'_{LL}=
\left( \begin{array}{ccc}
0.437972  & -0.897698 &  -0.9773193\\
 -0.897698 & -0.443296  & -0.230068\\
 -0.9773193  & -0.230068 & -0.005324  
\end{array}
\right)m_{0}
\end{equation}
and for Type I(A) in Eq.(6) we have 
\begin{equation}
m'_{LL}=
\left( \begin{array}{ccc}
0.32668  & -0.74163 &  -0.57122\\
 -0.74163 & 0.17014  & -0.64185\\
 -0.57122  & -0.64185 & 0.50306  
\end{array}
\right)m_{0}
\end{equation}
It is interesting to note that both $m'_{LL}$ in the above Eqs.(18) and (19)
have now acquired  relatively larger  $|m_{ee}|$ consistent with the double beta decay result[1].

A few comments on the stability condition under radiative 
corrections are in order. The nearly degenerate mass matrices
$m_{LL}$ in Eqs.(6),(8) and (10) are found to be unstable under 
radiative correction in minimal supersymmetric standard model(MSSM).
The inverted hierarchical mass matrix with the same mass eigenvalues
given in Eq.(12) is also found to be unstable under radiative
correction. However, the inverted hierarchical mass matrix given in
Eq.(14) with opposite sign mass eigenvalues, is stable under
radiative correction [12,14]. The radiative stability of neutrino mass texture for nearly 
degenerate eigenvalues remains an outstanding problem at the moment[15,16].  

\section{Summary}
 In summary, we generate the textures of the nearly degenerate 
as well as the inverted hierarchical left-handed Majorana neutrino 
mass matrices from the see-saw formula using the diagonal form of 
the Dirac mass matrix and non-diagonal form of the right-handed Majorana
neutrino mass matrix. This is a continuation of our earlier work where bimaximal mixings are generated 
from the texture of $M_{RR}$ in case of hierarchical and inverted hierarchical models.
The predictions on lepton mixing angles 
$sin^{2}2\theta_{12}\approx 0.67$, $sin^{2}2\theta_{23}\approx 1.0$ and
$|V_{e3}|\approx 0$ are in excellent agreement with the experimental 
values in all cases except for types I(A) and II(B).
We also get good  predictions for   $\Delta m^{2}_{12}$ 
and $\Delta m^{2}_{23}$ which are necessary for the 
$0\nu \beta \beta$ decays, LMA MSW solar oscillation
and atmospheric oscillation data. In all cases the masses of the right-handed Majorana
neutrinos are  above the weak scale. With the inclusion of the contribution from the diagonalisation 
of the charged lepton matrix having a special structure in 1-2 block, we improve the prediction on solar mixings to the right order for types I(A) and II(B).\\
 Though the present work is a model independent analysis without using any underlying symmetry, it would serve as a useful guide to building models under the framework of grand unified theories with extended flavour symmetry. In short the present analysis explores the possible  origin of the bimaximal neutrino mixings from the texture of right-handed Majorana mass matrices.

\end{document}